# Compositional and *in Vitro* Evaluation of Nonwoven Type I Collagen/Poly-dl-lactic Acid Scaffolds for Bone Regeneration

Xiangchen Qiao [1,3], Stephen J. Russell [2], Xuebin Yang [1], Giuseppe Tronci [1,2] and David J. Wood [1,]*

[1] Biomaterials and Tissue Engineering Research Group, School of Dentistry, University of Leeds, Leeds LS2 9LU, UK; E-Mails: hflionell6318@hotmail.com (X.Q.); x.b.yang@leeds.ac.uk (X.Y.); g.tronci@leeds.ac.uk (G.T.)

[2] Nonwovens Research Group, Centre for Technical Textiles, University of Leeds, Leeds LS2 9JT, UK; E-Mail: s.j.russell@leeds.ac.uk (S.R.)

[3] State Key Laboratory of Oral Diseases, Sichuan University, Chengdu 610041, China

* Author to whom correspondence should be addressed; E-Mail: d.j.wood@leeds.ac.uk (D.W.) Tel.: +44-113-343-6192; Fax: +44-113-343-6548.

## Abstract

Poly-dl-lactic acid (PDLLA) was blended with type I collagen to attempt to overcome the instantaneous gelation of electrospun collagen scaffolds in biological environments. Scaffolds based on blends of type I collagen and PDLLA were investigated for material stability in cell culture conditions (37 °C; 5% $CO_2$) in which post-electrospinning glutaraldehyde crosslinking was also applied. The resulting wet-stable webs were cultured with bone marrow stromal cells (HBMSC) for five weeks. Scanning electron microscopy (SEM), confocal laser scanning microscopy (CLSM), Fourier transform infra-red spectroscopy (FTIR) and biochemical assays were used to characterise the scaffolds and the consequent cell-scaffold constructs. To investigate any electrospinning-induced denaturation of collagen, identical PDLLA/collagen and PDLLA/gelatine blends were electrospun and their potential to promote osteogenic differentiation investigated. PDLLA/collagen blends with w/w ratios of 40/60, 60/40 and 80/20 resulted in satisfactory wet stabilities in a humid environment, although chemical crosslinking was essential to ensure long term material cell culture. Scaffolds of PDLLA/collagen at a 60:40 weight ratio provided the greatest stability over a five-week culture period. The PDLLA/collagen scaffolds promoted greater cell proliferation and osteogenic differentiation compared to HMBSCs seeded on the corresponding PDLLA/gelatine scaffolds, suggesting that any electrospinning-induced collagen denaturation did not affect material biofunctionality within 5 weeks *in vitro*.

## 1. Introduction

Bone repair is a subject of intensive investigation in human health care in light of the ageing population, and bone grafting is still the clinical gold standard for the restoration of lost bone structure and function from either traumatic or non-traumatic destruction [1,2]. Bone tissue engineering is becoming a promising medical route for large defect repair; a major issue is to design a biomimetic matrix as an agent for delivering cells or carrying tissue-inducing factors. An ideal biomimetic matrix should precisely restore the composition, architecture, and organized pattern of natural tissue [2].

Type I collagen is the most abundant protein in bone. It is a triple helix of three polypeptides [3,4]. The collagen fibrils aggregate laterally and longitudinally together to give the material its characteristic hierarchical structure [5–7]. The tensile strength of collagen is a result of intermolecular covalent crosslinks between the telopeptides and the helical regions of neighbouring collagen molecules. Moreover, natural collagen is characterised by its crystallinity, triple-helix structure and protease resistance [8].

Electrospinning of collagen has been widely applied as a one-step process for the formation of fibrous materials mimicking tissue structure from the molecular to the microscopic level. To ensure the formation of defined and porous fibrous architecture, growing attention has been paid to the selection of an appropriate electrospinning solvent for collagen. Since being first reported by Matthews *et al.*, most studies have adopted the use of fluoroalcohol, (1,1,1,3,3,3)-hexafluoro-2-propanol (HFIP) as the preferred solvent for collagen electrospinning [9–13]. However, studies have indicated that electrospun collagen is unstable in water [14–17] leading some to suggest that this material is effectively denatured collagen, *i.e.*, gelatine [8]. To overcome the issue of stability, co-electrospinning of collagen with either synthetic or natural polymers, such as PCL [18,19], polyurethane [20], silk fibroin [21], chitosan [22], and poly-l-lactide (PLLA) [23], has been widely adopted.

PDLLA is an amorphous polymer formed via polymerization of a racemic mixture of L- and D-lactides [24,25]. The precise composition of the polymer determines its mechanical properties and hydrolysis characteristics; in comparison to the semicrystalline PLA homopolymer, PDLLA displays faster degradation [26-30], due to the rapid access of water in the amorphous material and the hydrolytic cleavage of polymer ester bonds.

Electrospinning of PDLLA can induce changes in the polymer wettability, whilst a range of fibre morphologies can be obtained. On the one hand, the high surface tension at the air-polymer interface, the whipping and rapid solvent evaporation and solidification can lead to rough fibres with increased hydrophobicity [31]. On the other hand, in electrospun PDLLA fibres, it has also been claimed that chemical groups are attached to the fibre surface as a result of the high voltage used in electrospinning, resulting in an increased water contact angle [30].

For the present study, in light of PDLLA hydrophobicity [27,29,32,33], it was hypothesised that co-electrospinning of PDLLA with type I collagen could enhance the wet stability of the resulting scaffolds, thereby enabling the use of this material for tissue engineering applications [29], The addition of a hydrophobic phase in the electrospun material was expected to reduce the collagen swelling with the aqueous media, thereby enhancing the material wet-stability.

A pre-requisite for the electrospinning of a mixed polymer solution of type I collagen and PDLLA is the selection of an appropriate co-solvent system. Relatively few solvents are available for the electrospinning of collagen [9,34-36], and none of these are conventionally selected for the electrospinning of PDLLA [27,29,37,38].

Therefore, in order to enable an evaluation of the suitability of PDLLA to stabilise electrospun fibres containing collagen, an objective of the present study was to identify a suitable co-solvent system for PDLLA/collagen. Subsequently, electrospun PDLLA/collagen webs could be produced with different wettabilities, degradation behaviour [39], and cell growth potential [40]. A second objective was therefore to conduct *in vitro* biological characterisation of the new scaffolds, which was undertaken with human bone marrow stromal cells. Finally, a third objective was to identify any differences in the biofunctionality of electrospun PDLLA/collagen with respect to PDLLA/gelatine scaffolds, in light of any electrospinning-induced alteration of the collagen native conformation.

## 2. Results and Discussion

### 2.1. Physical Wet-Stability of PDLLA/Collagen Scaffolds

As shown in Figure 1A–D, all electrospun samples were composed of randomly oriented fibres that formed an interconnected porous network. The ratio of PDLLA to collagen was found to influence fibre diameter. As shown in Table 1, under fixed spinning conditions, both the mean fibre diameter and average porosity progressively decreased with increasing PDLLA content.

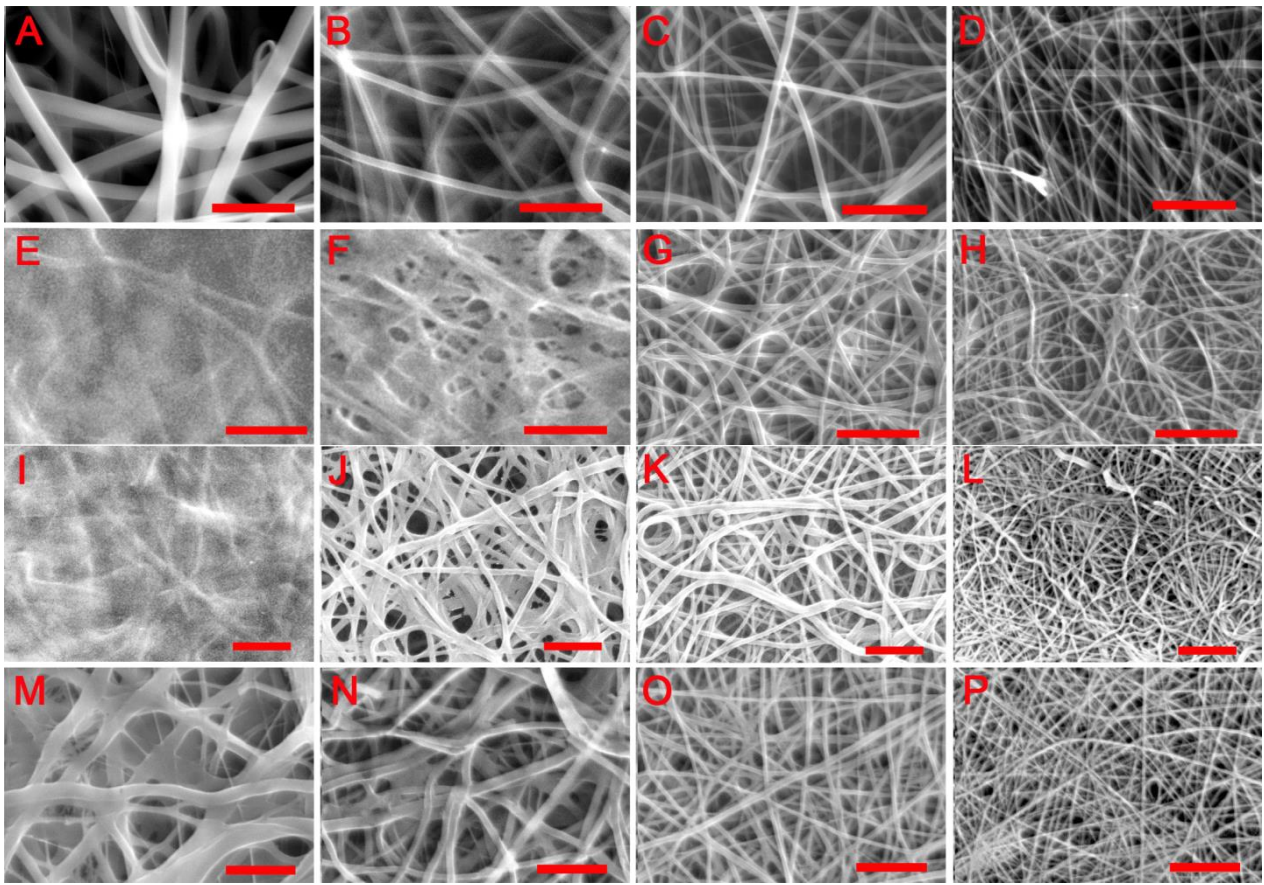

**Figure 1.** SEM images of electrospun PDLLA/collagen scaffolds. (**A**), (**B**), (**C**) and (**D**) are PDL20, PDL40, PDL60 and PDL80 as-spun scaffolds, respectively; (**E**), (**F**), (**G**) and (**H**) are PDL20, PDL40, PDL60 and PDL80 scaffolds after water immersion, respectively, water immersion was conducted in deionised water overnight in room temperature; (**I**), (**J**), (**K**) and (**L**) are PDL20, PDL40, PDL60 and PDL80 scaffolds after incubation in distilled water overnight (37 °C, 5% $CO_2$ and 95% relative humidity), respectively; (**M**), (**N**), (**O**) and (**P**) show the impact of GTA crosslinking on the stability of PDL20, PDL40, PDL60 and PDL80 scaffolds, respectively; crosslinking was performed using GTA vapour (25%, w/v) at room temperature, and the resultant scaffolds were incubated in distilled water overnight (37 °C, 5% $CO_2$ and 95% relative humidity) before imaging. Scale bar = 10 μm.

Figure 1E–H indicates the morphological changes in the scaffolds resulting from immersion in water overnight at room temperature. Evidence of fibre swelling was observed in some samples

leading to decreased scaffold pore size. The morphological changes were dependent on the PDLLA:collagen ratio. A low PDLLA:collagen ratio led to apparent merging of adjacent fibres resulting in partial loss of the original fibrous architecture in PDL20 (Figure 1E) and PDL40 (Figure 1F). For PDL60 (Figure 1G) and PDL80 (Figure 1H), fibre swelling was observed (Table 1) but the fibrous structure was clearly retained. Immersion also induced a decrease in scaffold porosity for all compositions, as shown in Table 1.

**Table 1.** Dimension properties of electrospun PDLLA/collagen scaffolds.

| Composite Scaffolds | Fibre Diameter (nm) | Porosity % | Fibre Diameter (nm) | Porosity % |
|---|---|---|---|---|
| | *As-spun* | | *After wet immersion* | |
| PDL20 | 1686 ± 55.5 | 70.1 ± 2.43 | N/A | 0.35 ± 0.23 |
| PDL40 | 1014 ± 83.4 | 68.5 ± 1.16 | N/A | 9.1 ± 2.11 |
| PDL60 | 668 ± 23.4 | 65.4 ± 1.42 | 701 ± 19.9 | 54.9 ± 1.53 |
| PDL80 | 330 ± 13.2 | 59.5 ± 8.71 | 384 ± 10.2 | 53.7 ± 2.76 |
| | *Incubation/no GTA* | | *Incubation/GTA* | |
| PDL20 | N/A | 0.6 ± 0.02 | 1692 ± 71.6 | 55.36 ± 3.35 |
| PDL40 | 1303 ± 48.5 | 59.8 ± 3.08 | 1052 ± 35.3 | 62.8 ± 3.97 |
| PDL60 | 770 ± 23.3 | 55.7 ± 4.2 | 677 ± 47.4 | 63.3 ± 1.05 |
| PDL80 | 472 ± 15.5 | 54.5 ± 2.89 | 336 ± 39.7 | 51.7 ± 2.05 |

Similar morphological changes were observed when scaffolds were incubated at 37 °C as opposed to room temperature (Figure 1I–L), but the fibrous structure of PDL40 scaffolds (Figure 1J) was better retained than those that were subject to wet immersion at room temperature (Figure 1F).

A summary of mean fibre diameters and porosities of the scaffolds after spinning, after immersion in water at room temperature, or after incubation at 37 °C is presented in Figure 2A,B, together with statistical analysis of the data. In the as-spun scaffolds, fibre diameter decreased with increasing PDLLA content. Water immersion at room temperature resulted in loss of fibrous structure in the PDL20 and PDL40 scaffolds so that an accurate estimation of mean fibre diameter could not be obtained (Figure 2A). The same was true for the PDL 20 scaffolds after incubation at 37 °C, but was not observed for the PDL 40 scaffolds after incubation. For all compositions, water immersion at room temperature or incubation (37 °C) inevitably led to significant fibre swelling ($p < 0.001$). Following either immersion or incubation, a decrease in porosity was observed for all scaffolds (although this was not found to be statistically significant for PDL80, probably due to the relatively high standard deviation for the mean porosity of the as-spun samples). The porosities of the PDL20 scaffolds after water immersion at room temperature and 37 °C were negligible (Figure 2B). The same was true for PDL40 scaffold after water immersion at room temperature; however, the retention of porosity in the PDL40 sample was significantly improved after incubation at 37 °C overnight compared to room temperature water immersion ($p < 0.001$) (Figure 2B).

Regardless of the PDLLA to collagen ratio, although most evident for PDL20, fibre swelling at 37 °C was markedly reduced following GTA vapour crosslinking (Figure 1M–P and Table 1) compared to the non-treated groups (Figure 1I–L and Table 1). However, merging of fibres was observed in PDL20 and PDL40 scaffolds (Figure 1M,N, respectively) and partial collapse of the scaffold layers in the PDL20 scaffolds was noted (Figure 1M).

Statistically, it was found that after GTA vapour crosslinking (data obtained after water incubation at 37 °C), the fibre diameter for all the scaffolds ($p < 0.001$) was significantly reduced (Figure 2C). For the PDL20 scaffold, its porosity and mean fibre diameter was significantly improved ($p < 0.01$) after GTA vapour crosslinking (water incubation at 37 °C) (Figure 2 C,D); however, the scaffold had noticeably collapsed.

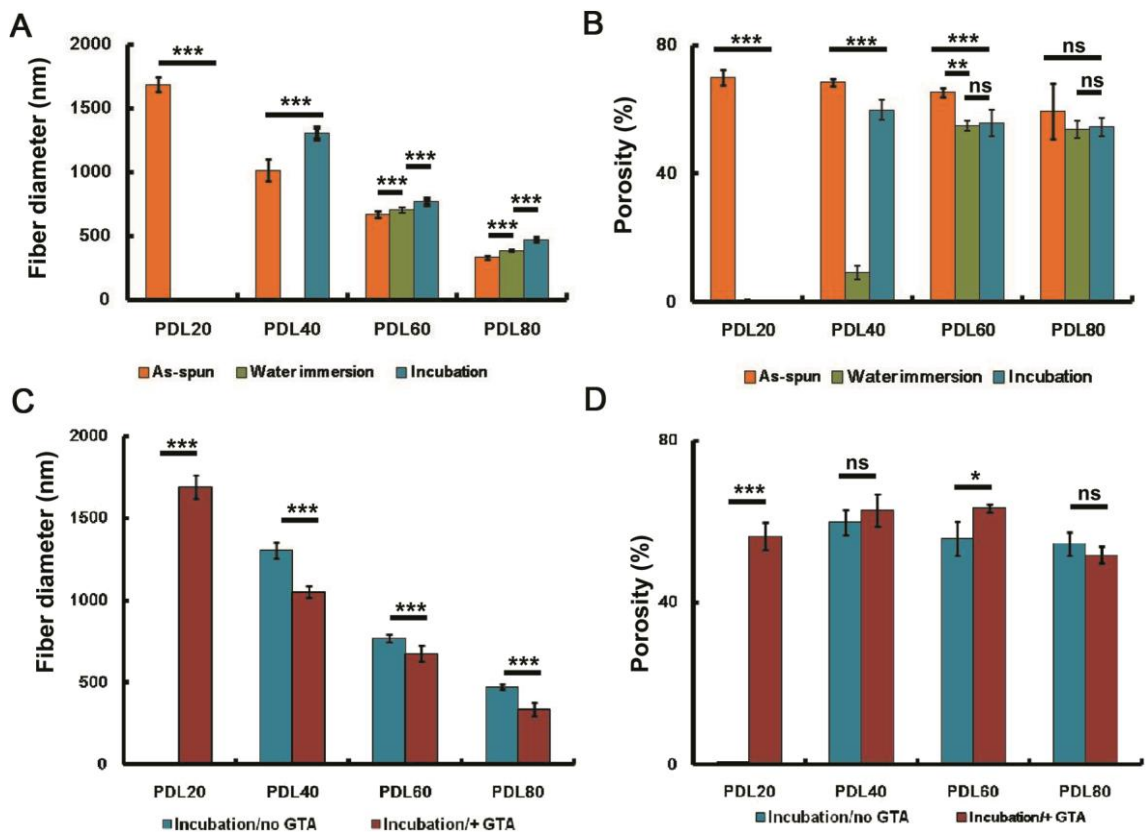

**Figure 2.** Influence of the treatments on the mean fibre diameter and porosity of PDLLA/collagen scaffolds. (**A**,**B**) show the morphological changes of scaffolds after water immersion (room temperature) and incubation (37 °C); (**C**,**D**) compare the difference of morphological changes with or without GTA crosslinking. (A) and (C) are mean fibre diameter; (B) and (D) are scaffold porosity. The error bars represent mean ± SD (sample size $n = 100$ for diameter, $n = 3$ for porosity). For porosity, data was derived after converting SEM images into binary images, porosity ($\varphi$) was given as (1-white pixels/total pixels) × 100% for each binary images. White pixels = fibres, total pixels = sum of material and voids. One-way ANOVA with Bonferroni post-test was performed, ns $p > 0.05$, **$p < 0.01$, *$p < 0.05$, ***$p < 0.001$.

Figure 3 shows the FTIR spectra of the electrospun PDLLA/collagen fibre materials. As the PDLLA content increased, peaks corresponding to PDLLA at 1748 cm$^{-1}$ and in the region 1300 to 1100 cm$^{-1}$ gradually increased, which is consistent with C=O stretching and C–O–C stretching or stretching vibration of the C–CH$_3$ group at 1039 cm$^{-1}$. It is known that amide I and II bands of collagen are present in the spectral range 1500 to 1700 cm$^{-1}$ [41-43]. As the collagen content in the scaffold increases, the amide I band, which appears at about 1650 cm$^{-1}$, gradually increases as a result of the C=O stretching vibrations coupled to N–H bending vibrations.

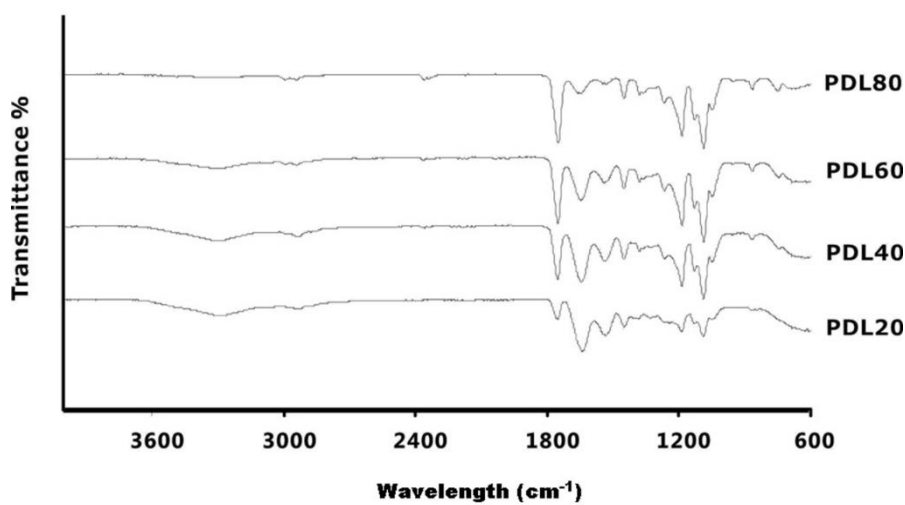

**Figure 3.** FTIR spectra of electrospun PDLLA/collagen scaffolds. FTIR spectra for PDL20, PDL40, PDL60 and PDL80. Corresponding to the increase of PDLLA contents in the blends, peaks at 1748 cm$^{-1}$ and in the region of 1300 to 1100 cm$^{-1}$ increase, while peaks in the region of 1500 to 1700 cm$^{-1}$ and at 1650 cm$^{-1}$ decrease.

*2.2. Cell Culture on PDLLA/Collagen Scaffolds*

Given the wet instability of the PDL20 scaffolds and the relatively poor retention of their fibrous structure, HBMSCs were cultured only on the GTA-vapour crosslinked PDL40, PDL60 and PDL80 scaffolds. The initial (16 h) cell attachment, together with 4-week proliferation and differentiation as well as morphological changes in the scaffolds, were determined. As shown in the Figure 4, attachment of HBMSCs on the scaffolds after 16 h of cell culture was analysed statistically. Following labelling with CMFDA and culturing overnight, attached cells were imaged on all selected scaffolds. The number of cells attracted to PDL40 (Figure 4C and the corresponding column) was significantly higher than the number of cells on the PDL60 (Figure 4B and the corresponding column) and PDL80 scaffolds (Figure 4A and the corresponding column). Cells on PDL40 were closely packed and spread over the solid surface of the fibres, while spherical cells were observed on both the PDL60 and PDL80 scaffolds.

The scaffold-cell constructs were observed using SEM after four weeks in either basal or osteogenic media as shown in Figure 5A–F. Figure 5A–C represent PDL40, PDL60 and PDL80 respectively, in basal medium. The underlying scaffold materials were visible for all of the cell-scaffold constructs from basal culture. This is in contrast to the cells cultured in osteogenic medium, where confluent cell matrix layers were observed on the surface of scaffolds (Figure 5D–F), indicating a higher density of cells.

The macroscopic appearances of scaffold-cell constructs were also observed using light microscopy after 4 weeks in either basal or osteogenic media. Macroscopically, a massive adverse change in sample geometry was observed in the PDL80 scaffolds, compared to PDL40 and PDL60, with the samples appearing to have shrunk/degraded and folded in on themselves.

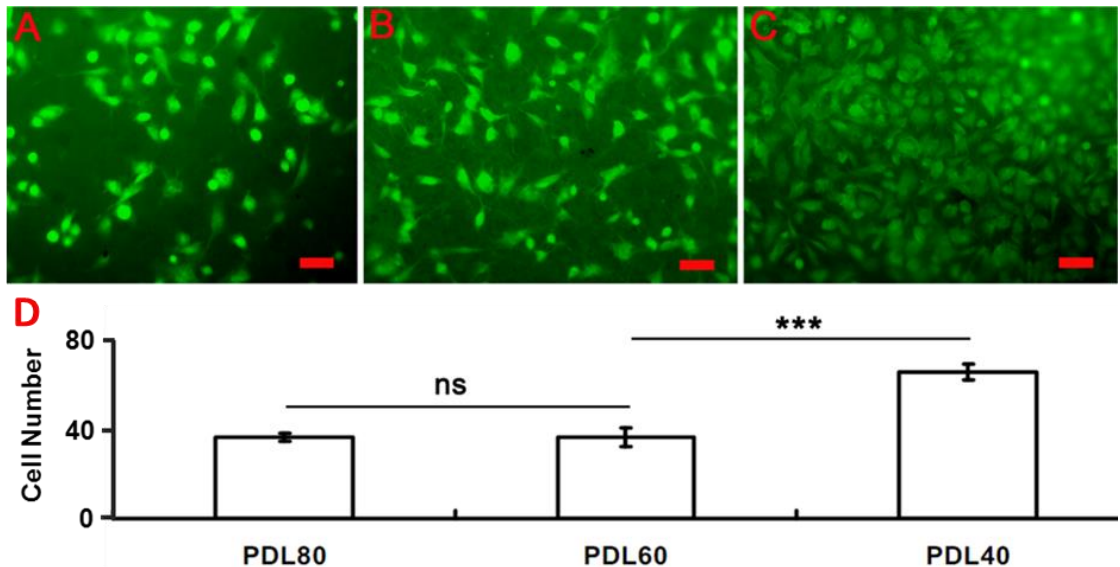

**Figure 4.** Attachment of HBMSCs on the PDLLA/collagen electrospun scaffolds. HBMSC attachment on (**A**) PDL80, (**B**) PDL60 and (**C**) PDL40 was determined. Cells (5 × 10⁴) were stained with CMFDA before seeding onto scaffolds (surface area = 3.8 mm²). Fluorescence microscope images were taken after 16 h for each scaffold ($n = 3$). (**D**) Cell numbers were determined and given as mean ± SD ($n = 3$), a two-tail, unpaired t-test was performed for comparison, ns $p > 0.05$, *$p < 0.05$, **$p < 0.01$. Scale bar = 20μm.

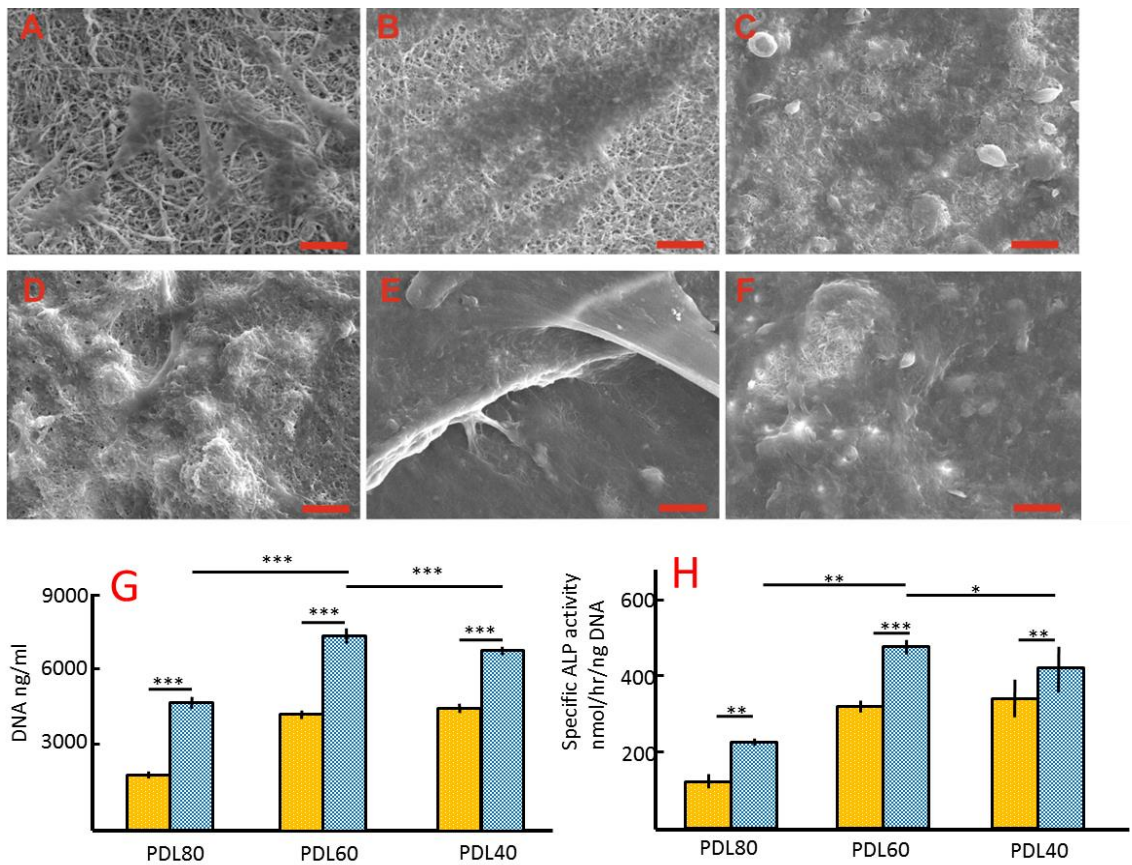

**Figure 5.** *In vitro* studies after 4-week basal and osteogenic culture. (**A**–**F**) are the SEM images of cell-PDLLA/collagen constructs after a 4-week culture. Isolated cells were observed on the basal culture groups (**A**–**C**), while a confluent cell layer was observed on osteogenic culture groups (**D**–**F**). Scale bar = 100 μm. (**G**) and (**H**) show the DNA quantity and specific ALP activities of HBMSCs on the scaffold after a 4-week

culture. The error bars represent mean ± SD ($n = 3$), on-way ANOVA with Bonferroni post-test was performed, ns $p > 0.05$, *$p > 0.05$, **$p < 0.01$, ***$p < 0.001$. Yellow bars represent basal media, blue bars represent osteogenic media.

The proliferation and osteogenic differentiation of cells on the PDLLA/collagen scaffolds were determined after four weeks of culture in basal and osteogenic culture media. DNA content was used to compare the proliferation of HBMSCs on the PDLLA/collagen scaffolds. Variations of DNA quantity are given in Figure 5G. Statistically, osteogenic culture significantly increased the DNA content compared to the basal medium for all scaffolds ($p < 0.001$). There was no significant difference in the DNA quantities between PDL40 and PDL60 basal culture groups ($p > 0.05$), but the quantity of DNA in the PDL60 osteogenic groups was significantly higher than in the other osteogenic groups ($p < 0.001$).

In the basal medium culture group, PDL60 and PDL40 scaffolds enhanced the HBMSC's ALPSA compared to that in the PDL80 sample ($p < 0.001$). However, there was no significant difference between the PDL 60 and PDL40 samples ($p > 0.05$) (Figure 5H). In osteogenic culture conditions, the PDL60 group appeared to give the highest ALPSA compared to the PDL80 ($p < 0.001$) and PDL 40 ($p < 0.05$) samples (Figure 5H). Comparing the two different culture conditions for the same scaffold, ALPSA was observed to increase in osteogenic culture conditions ($p < 0.01$) (Figure 5H).

*2.3. Comparison of PDLLA/Collagen and PDLLA/Gelatine Scaffolds*

Two sets of scaffolds were co-electrospun from blends of either collagen/PDLLA or gelatin/PDLLA under the same conditions (examples are shown Figure 6A,B). For each composition, bead-free fibrous structures with well-defined fibre morphologies were obtained. Cell (HBMSCs) adhesion on the scaffolds was imaged using confocal microscopy and maximum projection images are shown in Figure 6C,D. A two-tail, unpaired t-test was used to analyse the differences in cell attachment between the two series of scaffolds. A significantly greater number of cells attached to the PDL60/Col scaffolds as compared to the PDL60/Gel scaffolds (Figure 6E). Moreover, after overnight culture, the cells on the PDL60/Gel were still spherical in shape (Figure 6D) whilst cell spreading was observed on the PDL60/Col scaffolds (Figure 6C).

Cell behaviour was assessed after 5 weeks of culture *in vitro* for both basal and osteogenic culture conditions. Figure 7A indicates that the DNA content of the PDL60/Col groups was significantly ($p < 0.001$) higher than in the PDL60/Gel groups. This was consistent with ALPSA (Figure 7B).

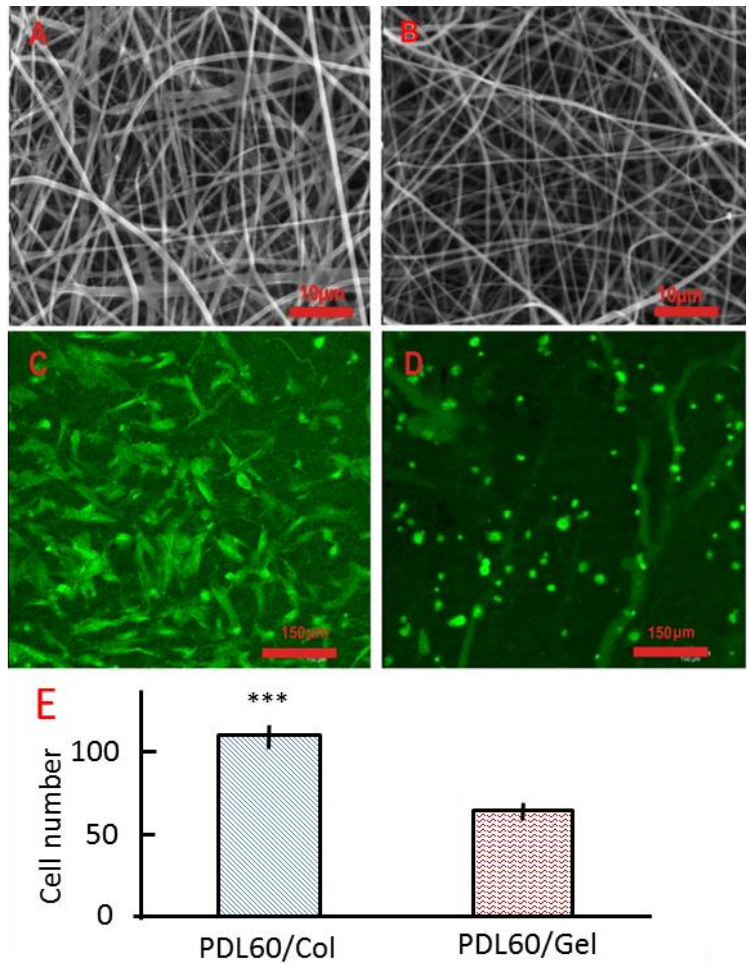

**Figure 6.** HBMSCs attachment on the PDL60/Col and PDL60/Gel electrospun scaffolds. (**A,B**) are SEM images of PDL60/Col scaffolds and PDL60/Gel scaffolds, respectively. (**C,D**) are maximum projection fluorescence images of PDL60/Col and PDL60/Gel cell-scaffold constructs, respectively. Images were collected using confocal microscopy with 20× water immersion objective. (**E**) Cell numbers were counted based on the confocal images and given as mean ± SD ($n = 3$), two-tail, unpaired t-test was performed, ***$p < 0.001$.

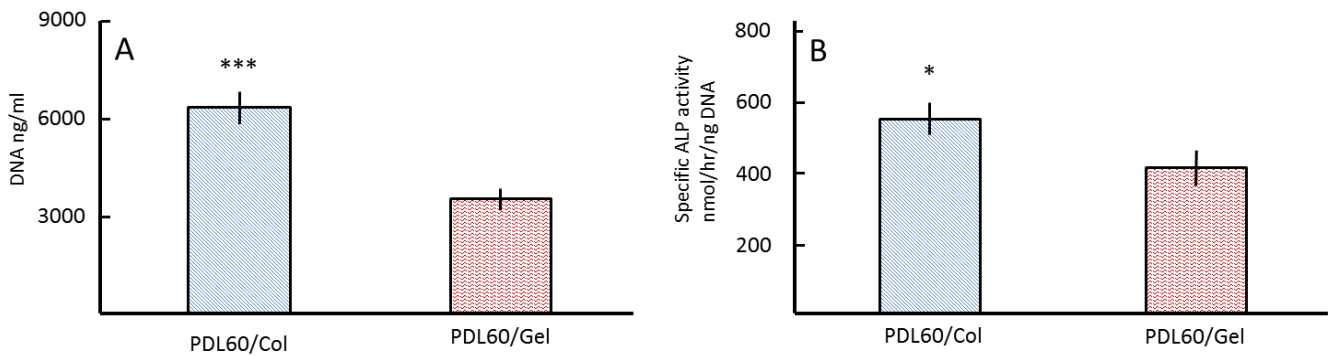

**Figure 7.** Proliferation and differentiation of HBMSCs on PDL60/Col and PDL60/Gel scaffolds. After seeding with HBMSCs for 5 weeks, the DNA content (**A**) and specific ALP activities (**B**) of HBMSCs were determined. The error bars represent means ± SD for sample size ($n = 3$); a two-tail, unpaired t-test was performed, *$p < 0.05$, ***$p < 0.001$.

*2.4. Discussion*

The chemical stability of fibres in terms of their wetting and dissolution behaviour influences the functional performance of tissue engineering scaffolds [31]. Collagen can be effectively electrospun to produce a fibrous scaffold, although the inherent compatibility of the as-spun fibres with water adversely affects the stability of the entire scaffold. The chemical compatibility with and degree of swelling in water is influenced by the molecular organisation of the polymer [44]. In contrast to regenerated collagen, PDLLA is a hydrophobic synthetic material that is known to reduce the water uptake of fibrous assemblies due to its surface energy [31]. The mixing of collagen with an insoluble, water-stable polymer such as PDLLA can be expected to affect the bulk properties of the resulting material, particularly if the two components are not phase-separated during electrospinning. The presence of both collagen components and PDLLA in the electrospun scaffolds was confirmed by FTIR analysis (Figure 3).

2.4.1. Physical Wet-Stability of PDLLA/Collagen Scaffolds

In the as-spun scaffolds, it was shown (Figure 1A–D) that increased PDLLA content results in a decreased mean fibre diameter. This reflects results described by Cui *et al*. [29] where addition of lower molecular weight polymers, resulted in the formation of thinner fibres. Any change in the solution viscosity as a result of changing polymer concentration can be expected to affect the rheology of the polymer stream between the spinneret tip and the collector, which will influence attenuation during electrospinning and the resultant fibre diameter. It was further suggested by Cui *et al*. [31] that a decrease in fibre diameter increased the surface roughness and entrapment of air in the scaffold, effectively reducing the hydrophilicity of the scaffold. In the present work, as expected, the wet stability of the fibres was found to improve as the PDLLA content increased (Figure 1E–H) while the mean fibre diameter decreased as a result of the lower collagen fraction. Additionally, the high voltage, the drag force and attenuation of the polymer stream during flight from the spinneret to the collector and rapid solvent evaporation during electrospinning are thought to influence molecular orientation in the fibre [31].

Fibre swelling, shrinkage and changes in the microstructure of the scaffolds were observed after water immersion and incubation (Figure 1E–L and Table 1). In addition to the inherent affinities of polar groups in the material leading to water adsorption in the amorphous regions and lateral swelling, polymer chain relaxation as a result of sorption may also be expected to affect fibre dimensions. Wet stability of the electrospun material could also be influenced by degradation or modification of the original molecular structure of the polymer components during the process. Dissolution of PDLLA in HFIP is rarely reported in relation to electrospinning since this solvent is normally associated with collagen. As a fluoroalcohol, HFIP could potentially degrade the side chains of polylactide chains via a catalytic degradation mechanism [32,45]. Although complete removal of residual solvent was rigorously conducted using the established vacuum drying (16 h) procedure at room temperature [27,39,46], the possibility of polymer degradation due to trace solvent residue cannot be entirely discounted.

The PDL40 scaffold exhibited unexpectedly improved wet stability after incubation in water, suggesting the possibility of molecular rearrangement during the incubation process. A similar trend has been observed in electrospun PLGA scaffolds, where a sudden increase in crystallinity was observed after buffered incubation at 37 °C [47]. Such an increase in the crystallinity would be expected to limit chain relaxation and the penetration of water molecules into the material. As previously mentioned, reconstitution of collagen fibrils *in vitro* is a temperature-dependent process [48-51]. The improved wet-stability of PDL40 after incubation may be due to the synergistic effect of collagen and PDLLA. The PDL40 sample may be close to the best composition for an electrospun PDLLA/collagen hybrid material because in this composition, the wetting and sorption behaviour of PDLLA is sufficient to reduce the diffusion of water in to the material, preserving the collagen phase. The reconstitution of crystalline collagen after incubation could act to restrain the relaxation of PDLLA chains and swelling of the fibre. In the PDL20 scaffold, the large morphological changes in the fibrous structure may be due to there being insufficient PDLLA to prevent water adsorption and diffusion between the polymer chains.

After crosslinking with GTA vapour (Figure 1M–P and Table 1), the fibre diameter was consistently smaller than that of untreated scaffolds, which is attributed to the improved stabilisation of the collagen phase of the fibre. Porosity can also decrease markedly in electrospun scaffolds due to wet-collapse of the planar structure, which exhibits relatively poor compressive strength. Interestingly, although the PDL20 showed open pores at the top surface, its porosity was determined to be close to zero using image analysis. This was because the threshold for imaging analysis included all layers of the collapsed GTA crosslinked (wetted) scaffold, as well as the top layer, effectively reducing the porosity close to zero. Interestingly, the porosity of PDL 40, PDL 60 and PDL 80 scaffolds was not hugely affected by GTA treatment (indeed there was no significant difference in porosity for the PDL40 and PDL80 samples), at least in the short time period covered in this study. This may point to the fact that GTA crosslinking may not be necessary for certain compositions to be stable, although caution should be taken in extrapolating this into longer time points.

2.4.2. Interactions of PDLLA/Collagen Scaffolds and HBMSC Cells

As aforementioned, the wet-stability of a scaffold is important to maintain a support structure for the adhesion and growth of cells. From the results (Figure 2A,B, and Table 1), morphological changes in the PDL20 scaffolds included a merging of adjacent fibre surfaces and a significant decrease of scaffold porosity. Accordingly, all the PDLLA/collagen scaffolds except PDL20 were evaluated biologically using HBMSCs. Comparing Figure 1I–L and Figure 1M–P, GTA crosslinking effectively reduced the tendency for collapse of the integral planar fibrous layers. Accordingly, due to their improved stability, GTA crosslinking was carried out for all scaffolds that were selected for cell-scaffold evaluations.

Figure 4 revealed that the initial adhesion of cells was significantly higher on PDL40 than on the other scaffolds. This suggests that the relatively high collagen content initially provided more favourable conditions for cell attachment. Additionally, the relatively low PDLLA content could be expected to enhance the wettability of the fibre surfaces [31] as well as promote increased cell proliferation [30]. Numerous works have discussed the importance of hydrophilicity in cell adhesion and proliferation as well as methods of modifying the surface wettability [29,31,38,39,52,53]. It was apparent that the selection of an 'optimum' PDLLA/collagen scaffold composition involves balancing the need to provide a highly water-stable construction with the need for fibre surface hydration.

After four weeks, there was no significant difference between the proliferation rate of cells on the PDL40 and PDL60 scaffolds (Figure 5G), while the osteogenic differentiation of cells on PDL40 scaffolds was significantly lower than the PDL60 group (Figure 5H). Moreover, proliferation and differentiation of cells on both the PDL40 and PDL60 groups were superior to the PDL80 scaffolds (Figure 5G–H). This may be regarded as a result of the compositional changes (PDL40) and the shrinkage (PDL80) of scaffolds. Although PDL80 showed no marked morphological change (after water immersion at room temperature and incubation at 37 °C with or without GTA crosslinking), the macroscopic shrinkage of PDL80 was severe compared to the other scaffolds. The loss and folding of scaffold material resulted in a reduction of available supporting area for cell attachment. Considering the low density of cells on the PDL80 scaffolds after 16 h, it was not surprising that both the proliferation and differentiation of cells was reduced in comparison with PDL40 and the PDL60 cell-scaffold constructs. For the PDL60 samples, although they attracted a lower number of cells at initial seeding compared to the PDL40 samples, both their proliferation and differentiation were superior to the PDL40 samples after four weeks of osteogenic culture. This may be a consequence of a compositional change to the PDL40 scaffold owing to the intrinsically soluble collagen component of the scaffold being washed out over time. Although initially it appeared that collagen and PDLLA had a synergistic effect on stabilising the structure of the PDL40 scaffold, certainly compared to PDL20, this was only an observation after 12 hours, and the protection (from PDLLA) against dissolution (both PDLLA and collagen) may be not have been sufficient for a longer culture period. Over a period of four weeks, the composition of the PDL40 scaffold may change due to the partial loss of the collagen component during each cycle of buffer replenishment. Consequently, it may be argued that a comparison of the proliferation and differentiation between the PDL40 and PDL60 cell-scaffold constructs was not appropriate. This is because the composition of the PDL40 scaffold would no longer be "PDL40", and the PDL60 scaffold may, after 4 weeks, actually contain more collagen than PDL40, as the higher PDLLA polymer content helps to retain a greater proportion of the original collagen component.

Regarding the PDL80 scaffold, despite containing the highest amount of PDLLA polymer, substantial shrinkage of the resultant cell-scaffold was observed after four weeks of culture. This may due to the solvent used during electrospinning. HFIP is not conventionally used for the electrospinning of PDLLA [27,29,37,38] and the effect of HFIP on PDLLA has not previously been reported. However, this result does reflect the reported work on electrospinning of another member of the PLA family, PLLA (300kDa), with collagen, which led to a "sticky" (gel-like) product after seven days of room-temperature incubation [23]. A possible reason for the reduced stability is the catalytic degradation of PDLLA by HFIP during preparation of the electrospinning solution [32,44] or the action of the solvent residue afterwards. However, for the PDL60 scaffold, its composition may be close to the optimal composition for combining these materials, where the collagen and PDLLA components synergistically control the wet-stability of the scaffold whilst retaining a fibrous hydrophilic architecture to encourage cell attachment and subsequent proliferation and differentiation.

2.4.3. Comparison of PDLLA/Collagen and PDLLA/Gelatine Electrospun Scaffolds

Electrospun scaffolds were produced from 60:40 PDLLA/collagen and 60:40 PDLLA/gelatine under identical conditions and evaluated biologically by seeding the scaffolds with HBMSCs. As is evident from Figure 6, PDLLA/collagen scaffolds provided a more favourable surface for cell adhesion than the PDLLA/gelatine scaffolds and the differences were statistically significant. These results suggest that despite the denaturation of collagen during its dissolution and processing in electrospinning, the electrospun product cannot be considered to be identical to gelatine. This contradicts the hypothesis of Zeugolis *et al*. [8] who suggested that electrospinning of collagen results in materials that are analogous to gelatine and as such, collagen electrospinning simply represented a rather expensive means of producing gelatine scaffolds.

It should be noted that there are differences between the denaturation of collagen in electrospinning and the preparation of gelatine. Preparation of gelatine involves extensive scission of chemical bonds along the main polypeptide chains [54]. In contrast, the present results for electrospun collagen suggest that the polypeptides were left intact after the process. Accordingly, the PDL60/collagen scaffold can be considered to be a composite of PDLLA and partially denatured collagen, whereas the PDL60/gelatine scaffold comprised of PDLLA and scissored peptides with a range of molecular weights. This could provide further support to the explanation given for the unexpected behaviour of PDL40 after wetting and after incubation, as the intact polypeptide chains experienced suitable conditions for the reconstitution of crystalline collagen fibrils.

## 3. Experimental Section

PDLLA (57KDa, PURASORB) was provided as a gift by Purac (Gorinchem, The Netherlands). HFIP was purchased from VWR (Lutterworth, UK). Gelatine powder was purchased from BDH (Poole, UK); unless noted otherwise, all other chemicals and reagents were received from Sigma (Gillingham, UK). Distilled water was used in all experiments. MSCs were purchased from Lonza (Castleford, UK).

### 3.1. Collagen Isolation

Type I collagen was acid-extracted from rat tails as described elsewhere [55,56]. The frozen rat tails were thawed in 70% (v/v) ethanol for 20 min to sterilise them before individual tendons were pulled out of the tendon sheath, minced, and placed in dilute acetic acid (17.4 mM) under aseptic conditions. This viscous mixture was centrifuged at 15,000 ×g for 1 h. The supernatant was collected, the resultant crude collagen solution was mixed with 0.1 M NaOH in a 6:1 ratio in order to neutralise the acetic acid, followed by second centrifugation at 10,000 ×g for 45 min. After discarding the supernatant, an equal volume of fresh acetic acid (17.4 mM) was used to re-solubilise the collagen. The type I collagen solution was lyophilised and the resultant powder stored at 4 °C until required.

### 3.2. Preparation of PDLLA/Collagen Scaffolds

Different weight fractions of PDLLA were mixed with collagen in order to identify the influence of composition on resulting fibrous scaffold properties. PDLLA/collagen samples namely PDL20 (PDLLA:collagen = 20:80), PDL40 (PDLLA:collagen = 40:60), PDL60 (PDLLA:collagen = 60:40) and PDL80 (PDLLA:collagen = 80:20) were produced using the electrospinning technique. Briefly, collagen and PDLLA (total polymer concentration: 10 wt./vol.%) were dissolved in HFIP via magnetic stirring at room temperature. The mixed polymer solutions were then electrospun via a flat nose needle

(18 gauge), at 25 kV operating at a feed rate of 0.72 mL/h and a working distance of 120 mm, depositing the scaffold on an aluminium foil collector. All scaffolds were rigorously vacuum-dried overnight to remove extraneous solvent prior to further evaluation. The retention of fibrous structure, fibre morphology and fibre dimensions were determined by SEM analysis under an accelerating voltage of 25 kV. In order to maximise the wet stability of the electrospun PDLLA/collagen scaffolds, chemical GTA vapour crosslinking was performed as described elsewhere [21,57,58]. Scaffolds were crosslinked in the vapour of a 2.5% (w/v) GTA solution for 4 h at room temperature. Excess GTA was removed by a 0.1 M glycine aqueous solution (12 h) [10,21,57]. Results were used to guide the selection of PDLLA/collagen compositions for subsequent biological testing.

FTIR Analysis

The chemical composition and functional groups within fibres of each PDLLA/collagen scaffold were identified by FTIR (FTIR Spectrum BX, Perkin-Elmer, Seer Green, UK) based on 128 scans and a resolution of 4 cm$^{-1}$. A wave-number range of 4000 to 600 cm$^{-1}$ for attenuated total reflectance mode was employed.

*3.3. Physical Wet-Stability of Electrospun PDLLA/Collagen Scaffolds*

Wet stability was evaluated by observing any morphological changes within the scaffold or by changes to fibre diameter, porosity and scaffold dimensions. The electrospun scaffolds were punched into circular samples (diameter = 22.1 mm, thickness = 0.5 mm). All samples (with or without GTA crosslinking) were immersed in de-ionised water at room temperature (overnight) or incubated in de-ionised water at 37 °C (overnight). Samples were recovered in triplicate, rinsed with distilled water and vacuum dried overnight to prepare samples for analysis.

Samples were sputter coated with gold in an argon atmosphere. The fibre diameter was measured using digital image processing (Image-Pro PLUS Analysis software, version 4.5.1.22, Media Cybernetic Inc., Rockville, USA) of SEM images. Mean fibre dimensions ($n = 100$) and variation were obtained using SPSS (Windows version 16.0.1, SPSS Inc., Guildford, UK).

Surface porosity was measured via digital image processing according to the method of Ghasemi-Mobarakeh *et al*. [59]. Briefly, thresholded SEM images were converted into a binary form. The porosity ($\varphi$) in each binary image was calculated using the mean intensity of the images ($n = 6$), which distinguishes between the solid and air fractions of the scaffold:

$$\varphi = (1 - n/N) \times 100\%$$

where $n$ is the number of white pixels in the binary image (fibre material) and $N$ is the total number of pixels in the binary image (sum of fibre material and void space).

*3.4. Biological Characterisation of Electrospun PDLLA/Collagen Scaffolds*

The results of the physical wet stability testing of scaffolds guided the selection of suitable scaffold compositions to take forward for cell culture studies. Given that dimensional stability in aqueous medium is desirable for cell culture [29,30] only those scaffolds exhibiting excellent stability were selected for *in vitro* cell culture.

Scaffolds were punched into circular samples (diameter = 22.1 mm, thickness = 0.5 mm) to cover the working area (3.8 cm$^2$) of a 12-well plate. All scaffolds were sterilised using UV (254 nm) irradiation on both sides for 30 min prior to cell seeding.

Human bone marrow stromal cells (HBMSCs) (Lonza) were cultured in DMEM supplemented with 10% FCS at 37 °C, 5% $CO_2$. Media were changed every four days. Cells were passaged at 80% confluency up to a maximum of passage four. Cell attachment was determined by inverted fluorescence microscopy after labelling of cells with Cell Tracker Green™ (5-chloromethylfluorescin diacetate, CMFDA). Briefly, a cell suspension (100 μL, $5 \times 10^4$ cells) was carefully seeded onto the surface of each scaffold overnight ($n = 3$). Cell-scaffold constructs were subjected to 98% ethanol fixation before they were examined microscopically. For each sample, images ($n = 3$) were taken from the most representative areas.

Proliferation and differentiation of cells was determined by quantitatively measuring their total DNA content and alkaline phosphatase-specific activity (ALPSA). Briefly, cell suspensions (100 μL, $5 \times 10^4$ cells) were seeded on the scaffolds ($n = 3$). Basal medium or osteogenic medium was added depending on the individual culture conditions. Media were changed every four days for a total of four weeks. The resultant cell-scaffold constructs were rinsed with 1× phosphate-buffered saline, and stored at −80 °C until required. Cell lysis was performed in a 1.5 mL Eppendorf tube (Eppendorf International) with the addition of 500 μL, of 0.1% Triton X-100 in carbonate buffer (133 mM $Na_2CO_3$, 67 mM $NaHCO_3$ and pH 10.2). Constructs were subjected to 5 min sonication (20 to 50 kHz) in ice-water and freeze-thawed three times. The concentration of DNA and ALPSA was determined using a fluorescence micro-plate reader (485 to 538 nm) and spectrometer (410 nm) according to the manufacturers' instructions (Sigma).

The morphologies of the cell-scaffold constructs were observed by SEM. Briefly, after three rinses with PBS, the samples were fixed in 98% ethanol, followed by dehydration in absolute ethanol and vacuum drying in a desiccator. Dry constructs were sputter coated with gold and observed by SEM at an accelerating voltage of 25 kV.

*3.5. Comparison of PDLLA/Collagen and PDLLA/Gelatine Electrospun Scaffolds*

Briefly, PDLLA was mixed with collagen or gelatine at the weight ratio of 60:40 and samples were denoted as PDL60/Col and PDL60/Gel respectively. The resultant blends were dissolved in HFIP to prepare 10% (w/v) of total weight for each composition. The polymer solutions were electrospun using a flat nose needle (18 gauge), at 25 kV operating with a feed rate of 0.72 mL/h and a working distance of 120 mm to an aluminium foil collector. Resultant scaffolds were vacuum dried overnight. CMFDA labelled HBMSCs ($1 \times 10^5$) were seeded on the scaffolds ($n = 3$). After fixing with 98% alcohol, cell attachment was determined after overnight post-seeding via confocal microscopy (CLSM, Leica SP2 TCS, Milton Keynes, UK). For each sample, three images (maximum projection along the $z$-axis) were taken from the most representative areas and cells were counted based on these multiple images using a 20× water immersion objective. Proliferation and differentiation of cells on each scaffold was determined by measuring their DNA quantity and alkaline phosphatase (ALP) activity accordingly.

*3.6. Statistical Analysis*

Data were gathered at least in triplicate and expressed as mean standard deviation. Statistical analysis was conducted using Prism V4.0 (GraphPad Software Inc., San Diego, CA, USA). Where appropriate, two-tail unpaired t test and one-way ANOVA with Bonferroni post-tests analyses were used. Fibre diameters were compared by analysis of variance (SPSS for Windows version 16.0.1, SPSS Inc.,). Differences were considered statistically significant at $p < 0.05$.

## 4. Conclusions

Incorporating high molecular weight PDLLA polymer with type I collagen followed by GD crosslinking produced water stable electrospun fibrous scaffolds. A PDLLA/collagen scaffold composition of 60/40 (w/w) was shown to be the most promising candidate for long-term cell culture. Evidence of a synergistic effect between the collagen and PDLLA components in stabilizing the overall scaffold structure during incubation was also observed. A fibrous scaffold containing collagen promoted attachment of human bone marrow stromal cells and growth in a five-week osteogenic cell culture. With respect to cell attachment and spreading *in vitro*, whilst denaturation of collagen takes place during electrospinning, higher cell proliferation and expression of ALP are observed on electrospun collagen-PDLLA scaffolds as compared to electrospun gelatin-PDLLA materials. Accordingly, despite denaturation, electrospun collagen does not yield a tissue engineering scaffold with analogous properties to that of a gelatine scaffold.

**Author Contributions**

X.Q., S.R., D.W. and X.Y. conceived and designed the experiments; X.Q. performed the experiments; X.Q., S.R., G.T., D.W. and X.Y. analyzed the data; G.T. and D.W. wrote the paper.

**Conflicts of Interest**

The authors declare no conflict of interest.